\documentclass[twocolumn,superscriptaddress,floatfix,showpacs,nofootinbib]{revtex4-1}
\usepackage{graphics,amssymb,amsmath,epsfig,color}
\usepackage{graphicx}

\newcommand{\be}{\begin{equation}}
\newcommand{\ee}{\end{equation}}
\newcommand{\bea}{\begin{eqnarray}}
\newcommand{\eea}{\end{eqnarray}}
\newcommand{\Prob}{{\rm Prob}}

\begin{document}

\begin{center}

{\large \bf RANDOMNESS, INFORMATION, AND COMPLEXITY\footnote{This paper appeared first in {\it 
          Proceedings of the 5th Mexican School on Statistical Physics (EMFE 5)}, 
          Oaxtepec, M\'exico, 1989, ed. F. Ramos-G\'omez (World Scientific, Singapore 1991); 
          the present version has some errors 
          corrected and some footnotes added to account for recent developments.}}

\vskip .5cm

           Peter Grassberger 
  
 {\it Physics Department, University of Wuppertal D 5600 Wuppertal 1, Gauss-­Strasse 20}

\end{center}
\vglue .7cm

{\bf Abstract:}

{\small
       We review possible measures of complexity which might in particular be applicable to 
situations where the complexity seems to arise spontaneously. We point out that not all of them 
correspond to the intuitive (or ``naive") notion, and that one should not expect a unique 
observable of complexity. One of the main problems is to distinguish complex from disordered 
systems. This and the fact that complexity is closely related to information requires that
we also give a review of information measures. We finally concentrate on quantities which 
measure in some way or other the difficulty of classifying and forecasting sequences of discrete 
symbols, and study them in simple examples.}

\vglue 1.4cm

\section{INTRODUCTION}

     The nature which we observe around us is certainly very ``complex", and the main aim of 
science has always been to reduce this complexity by descriptions in terms of simpler laws. 
Biologists, social and cognitive scientists have felt this most, and in these sciences attempts 
to formalize the concept of complexity have a certain tradition.

     Physicists, on the other hand, have often been able to reduce the complexity of the situations 
they were confronted with, and maybe for that reason the study of complex behavior has not been 
pursued very much by physicists until very recently.

     This reduction of complexity in physical systems is on the one hand achieved by studying 
archetypical situations (``Gedanken experiments"). On the other hand, very often when a complete 
description of a system is infeasible (like in statistical mechanics) , one can go over to a 
statistical description which might then be drastically more simple.

    Thus, in the most complex situations studied traditionally in statistical mechanics, the 
large number of degrees of freedom can usually be reduced to few ``order parameters". This is 
called the ``slaving principle" by H. Haken [1]. Mathematically, it is related to central limit 
theorems of statistics and to center manifold theorems [2] of dynamical system theory. Central 
limit theorems say that for large systems with many independent degrees of freedom statistical 
descriptions in terms of global variables become independent of details. Center manifold theorems 
say that in cases with very different time scales (e.g., near a bifurcation point) the fast modes 
can be effectively eliminated, and it is the slow modes which dominate the main behavior.
   
    But there are extremely simple systems (like the cellular automaton called ``game of life" [3]) 
which can be programmed as universal computers. One can of course try a purely statistical 
description also in this case. But it can be very inappropriate. One can never be sure that some 
of the details not captured by such a description do not later turn out to have been essential.

    And there are dynamical systems like those describable by a quadratic map
\be
    x_{n+1} = a - x_n^2,\quad x_n \in [-a,a]\;, \quad a \in [0,2] 
\ee
where even the reduction to a single order parameter does not imply a simple behavior. The 
complexity of the latter can be seen in a number of ways. First of all, the behavior can depend 
very strongly on the parameter $a$: there is a set of positive measure on which the attractor 
is chaotic [4], but this is believed to be nowhere dense, while windows with periodic attractors 
are dense. Secondly, at the transitions from periodicity to aperiodicity there are (an infinite 
number of) ``Feigenbaum points" [5], each of which resembles a critical phenomenon. The richness
inherent in Eq.(1) becomes even more obvious if we let $x_n$ and the parameter $a$ be complex.
The resulting Julia and Mandelbrot sets [6] (see e.g. Fig.~1) have become famous for their 
intricate structure even among the general public.

\begin{figure}
\vglue -2.3cm
\includegraphics[width=0.5\textwidth]{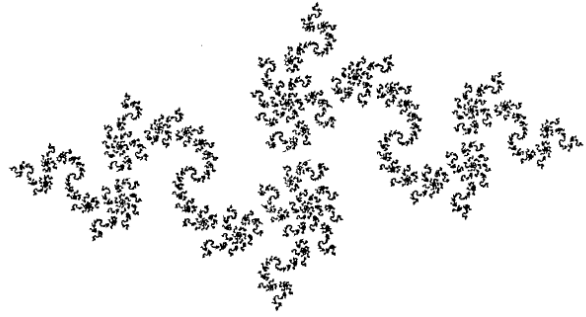}
\vglue -3.9cm
\caption{Julia set of the map $z' = z^2 - 0.86 - 0.25 i$.}
\end{figure}

Finally (and most importantly for us), the trajectories of Eq.(1) themselves can be very complex.
To be more specific, let us first discretize trajectories by defining [7] variables $s_n \in \{L,R,C\}$
with $s_n = L$ if $x_n<0$, $s_n = R$ if $x_n>0$, and $s_n = C$ if $x_n=0$. Similar ``symbolic dynamics"
can be defined also for other dynamical systems, with the mapping from the trajectories $(x_n; n\in Z)$
into the ``itineraries" $(s_n)$ being one-to-one for all starting points $x_0$. In general, one is indeed
satisfied if this mapping is 1-to-1 for {\it nearly} all sequences. In this case, one can drop the 
symbol ``$C$", and encode the trajectories of the quadratic map (of a large class of one-humped maps 
indeed) by binary $(R,L)$-sequences. The complexity of the map is reflected then in the fact that the 
itineraries obtained in this way show very specific structure, both ``grammatically" (i.e., 
disregarding probabilities) and probabilistically.

     Systems of similar complexity are e.g. the output of nonlinear electronic circuits, reversals 
of the earth's magnetic field, and patterns created by chemical reactions or by hydrodynamic 
turbulence. Beautiful examples of the latter are pictures of Jupiter, with numerous turbulent eddies 
surrounding the giant red spot.

     One characteristic common to all these instances is that the complexity is self-­generated in
the narrow sense that the formulation of the problem is translationally invariant, and the 
observed structure arises from a spontaneous breakdown of translational invariance (in the case of 
Jupiter, it is azimuthal rotation invariance which is broken). But the most interesting case
of self-generated complexity in a wider sense is presumably life itself.

     If we want to understand better the generation of complex behavior, we have at first a 
semantic problem: there does not seem to exist a universally accepted and formalized notion of what 
is ``complexity", though most of us certainly would agree intuitively that such a notion should exist.
As physicists used to work with precise concepts it should thus be one of our first aims to find 
such a precise notion. In the ideal case, it should be fully quantitative, i.e. there should be an 
observable and a prescription for measuring it.

     Actually, the situation is even worse: the most widely known concept of complexity of symbol 
sequences, the ``{\it algorithmic complexity}" [8,9], is actually a measure of information (following 
Ref.[l0], we shall thus call it below ``algorithmic information"). In the situations we are interested 
in, it is indeed closer to randomness than to complexity, although the similarities and differences 
are rather subtle. That information, randomness and complexity are closely related concepts should 
not be a surprise. But most of us feel very strongly that -- even if we cannot pin down this 
difference -- there is a crucial and all-­important difference between complex and merely random 
situations.

     For this reason, we shall in the next situation review information measures. In particular, we 
shall confront the Shannon information (which is a statistical measure and is indeed very closely 
related to thermodynamic entropy) [11] with purely algorithmic (i.e., non­probabilistic) measures,
the most prominent of which is algorithmic information itself. While the Shannon information can 
only deal with ensembles, and cannot strictly spoken attribute an information content to an 
individual message, the latter are designed to apply to single messages. Apart from very practical 
applications to data compression problems, the main application of algorithmic information is to 
sequences like the digit string of $\pi = 3.14159\ldots$, a (hopefully existing) proof of Fermat's 
last theorem, or the DNA sequence of Albert Einstein. These obviously are unique objects and should 
not be considered as just randomly chosen elements of some ensembles. Indeed, in these last examples 
the question of randomness does not arise, and thus information content and randomness cannot be 
the same. This is not true of the applications we are interested in. There, we just want an
observable which can help us in making this distinction.

     In sec.3, we shall come back to our central goal of finding a measure of complexity for these 
cases. In a search for such an observable, we shall make a list of required properties, and equipped 
with this we shall scan through the literature. We'll find several concepts which have been proposed, 
and all of which have advantages and drawbacks. We shall argue that indeed no unique measure of 
complexity should exist, but that different definitions can be very helpful in their appropriate 
places.

    Finally, in the last section, we shall apply these concepts to measure the complexity of some 
symbol sequences like those generated by the quadratic map or by simple cellular automata.

\section{ INFORMATION MEASURES}

\subsection{Shannon Information [11]}

          We consider a discrete random variable $X$ with outcomes $x_i , i = 1,\ldots N$. The 
probabilities $\Prob(X=x_i)$ are denoted as $P_i$.
They satisfy the constraints $0 \leq P_i \leq 1$ and $\sum_iP_i = 1$. The {\it entropy} or 
{\it uncertainty function} of $X$ is defined as
\be
   H(X) = -\sum_{i=1}^N P_i \log P_i 
\ee
(here and in the following, all logarithms are taken to base 2). It has the following properties 
which are easily checked:

 (i) $H(X) \geq 0$, and $H(X) = 0$ only if all $P_i = 0$ except for one single outcome $x_1$.

 (ii) For fixed $N$, the maximum of $H(X)$ is attained if all $P_i$ are equal. In this case, 
$H(X) = \log N$.

    These two properties show that $H$ is indeed a measure of uncertainty. More precisely, one can 
show that (the smallest integer $\geq$) $H(X)$ is just the average number of yes/no answers one 
needs to know in order to specify the precise value of $i$, provided one uses an optimal strategy. 
Thus, $H(X)$ can also be interpreted as an average information: it is the average information 
received by an observer, who observes the actual outcome of a realization of $X$. Notice that 
Eq.(2) is indeed of the form of an average value, namely that of the function $\log 1/P_i$.

    Equation (2) would not be unique if properties (i) and (ii) were all we would require from an 
information measure. Alternative ansaetze would be e.g. the {\it Reny\'i entropies}
\be
   H^{(q)}(X) = (1-q)^{-1} \log\left(\sum_{i=1}^N P_i^q\right)\;.  
\ee

What singles out the ansatz (2) is that with it, information can be given piecewise without loss, 
i.e. in some sense it becomes an additive quantity. What we mean by this is the following: Assume 
that our random variable $X$ is actually a pair of variables,
\be
    X = (Y, Z) \;,              \label{YZ}
\ee 
with the outcome of an experiment labeled by two indices $i$ and $k$,
\be
    x_{ik} = (y_i, z_k) \;.      \label{yz}
\ee
We denote by $p_k$ the probability $\Prob(Z=z_k)$ , and by $p^{(X)}_{i|k}$ the conditional probability 
$\Prob(Y=y_i | Z=z_k)$. Similarly, we denote by $H(Z)$ the entropy of $Z$, and by $H(Y|z_k)$ the 
entropy of $Y$ conditioned on $Z = z_k,\;\; H(Y|z_k)= -\sum_i p^{(X)}_{i|k}\log p^{(X)}_{i|k}$. From Eq.~(2)
we get then
\bea
   H(X) & & \equiv H(Y,Z) = H(Z) + \sum_k p_k H(Y|z_k)               \nonumber \\
        & & \equiv H(Z) + H(Y|Z)\;.   \label{H_cond}
\eea

The interpretation of this is obvious: in order to specify the outcome of $X$, we can first give the 
information needed to pin down the outcome of $Z$, and then we have to give the average information 
for $Y$, which might -- if $Y$ and $Z$ are correlated --­ depend on $Z$.
It is easy to check that Eq.~(\ref{H_cond}) would not hold for the Reny\'i entropies, unless $Y$ and
$Z$ were uncorrelated, $P_{ik} = p^{(X)}_i p_k$.

Let us now study the difference 
\be
   R(Y,Z) = H(Y)+H(Z)-H(Y,Z) \;.
\ee
It is obvious that $R(Y,Z) = 0$ if $Y$ and $Z$ are uncorrelated. The fact that in all other cases
$R(Y,Z) > 0$ is not so easily seen formally, but is evident from the interpretation of $H$ as an
uncertainty: If there are correlations between $Y$ and $Z$, then knowledge of the outcome of $Z$ can
only reduce the uncertainty about $Y$, and can never enhance it. For this reason, is called a 
{\it redundancy} or a {\it mutual information}. It is the average redundancy in the transmitted 
information, if the outcomes of $Y$ and of $Z$ are specified without making use of the correlations. 
Also, it is the average information which we learn on $Y$ by measuring $Z$, and vice versa.

    We shall mainly be interested in applying these concepts to symbol sequences. We assume that the 
sequences $\ldots s_i s_{i+1} s_{i+2} \ldots$ with the $s_i$ chosen from some finite ``alphabet" are 
distributed according to some translation invariant distribution. This means that, for any $n$
and $k$,
\be
   P(s_1s_2\ldots s_n) =P(s_{1+k}s_{2+k}\ldots s_{n+k}) \;,
\ee
where $P(s_1s_2\ldots s_n)$ is the probability that $S_k = s_k$ for $1 \leq k \leq n$, irrespective
of the symbols outside the window $[1...n]$. These ``block probabilities" satisfy the Kolmogorov 
consistency conditions 
\bea
    & P(s_2\ldots s_n)& =\sum_{s_1} P(s_1s_2\ldots s_n)\;,   \nonumber \\
    & P(s_1\ldots s_{n-1})& =\sum_{s_n} P(s_1s_2\ldots s_n).
\eea

If there exists some $N$ such that
\be
   P(s_{N+1}|s_0s_1\ldots s_N) = P(s_{N+1}|s_1\ldots s_N)\;,
\ee
we say that the distribution follows an $N$-­th order Markov process.

    Extending the definitions of entropy to the $n$-tuple random variable $S = (S_1 \ldots S_n)$, we 
get the {\it block entropy}
\be
   H_n = - \sum_{s_1 \ldots s_n} P(s_1s_2\ldots s_n) \log P(s_1s_2\ldots s_n)\;.  \label{Hn}
\ee
   
    Assume now a sequence is to be described symbol after symbol in a sequential way. If this 
description is to be without redundancy, then the average length of the description (or {\it encoding}) 
per symbol is given by
\be
   h = \lim_{n\to\infty} h_n\;,               \label{h}
\ee
\be
   h_n = H_{n+1}-H_n\;.                      \label{hn}
\ee
The interpretation of $h_n$ is as the information needed to specify the $(n+1)$-st symbol, provided 
all $n$ previous ones are known. The limit in Eq.~(\ref{h}) means that we might have to make use of 
arbitrarily long correlations if we want to make the most compact encoding.

    In the case of an $N$-th order Markov process, the limit in Eq.~(\ref{h}) is reached at $n=N$, 
i.e. $h_n = h$ for all $n\geq N$. Otherwise, the limit in Eq.~(\ref{h}) is reached asymptotically 
from above, since the $h_n$ are non­increasing,
\be
   h_{n+1} \leq h_n \;.
\ee
This is clear from the interpretation of $h_n$ as average conditional information: Knowing more 
symbols in the past can only reduce the uncertainty about the next symbol, but not increase it.

    Following Shannon [11], $h$ is called the entropy of the source emitting the sequence, or 
simply the entropy of the sequence. If the sequence is itself an encoding of a smooth dynamical 
process (as the $L,R$ symbol sequences of the quadratic map mentioned in the introduction), then 
$h$ is called the {\it Kolmogorov-Sinai} or {\it metric entropy} of the dynamical process.
Before leaving this subsection, I should make three remarks:

(a) The name ``entropy" is justified by the fact that thermodynamic entropy is just the Shannon 
uncertainty (up to a constant factor equal to Boltzmann's constant $k_B$) of the (micro-­)state 
distribution, provided the macrostate is given.

(b) Assume that we have a random variable $X$ with distribution
$\Prob(X=x_i) = P_i$. But from some previous observation (or from a priori considerations) we have 
come to the erroneous conclusion that the distribution is $P_i'$, and we thus use the $P_i'$ for 
encoding further realizations of $X$. The average code length needed to describe each realization 
is then $\sum_i P_i \log 1/P_i'$, while it would have been $\sum_i P_i \log 1/P_i$, had we used the 
optimal encoding. The difference
\be
   K = \sum_i P_i \log\frac{P_i}{P_i'}
\ee
is called the Kullback-­Leibler (or relative) entropy. It is obviously non-­negative, and zero only 
if $P = P'$ -- a result which can also be derived formally.

(c) In empirical entropy estimates, one has to estimate the probabilities $P_i$ from the frequencies 
of occurrence, $P_i \approx M_i /M$. For small values of $M_i$, this gives systematic errors due to 
the nonlinearity of the logarithm in Eq.~(2). These errors can be
eliminated to leading orders in $M_i$ by replacing Eq.~(2) by [12] \footnote{Note added in reprinting:
Ref. [12] is now obsolete and should be replaced by P. Grassberger,  arXiv:physics/0307138 (2003).}
\be
   H(X) \approx \sum_{i=1}^N \frac{M_i}{M} \left[ \log M - \psi(M_i) - \frac{(-1)^{M_i}}{(M_i+1)M_i}\right]
\ee
where $\psi(x) = d\log \Gamma(x)/dx$. Alternative methods for estimating 
entropies of symbol sequences are discussed in Refs. [13,14].

\subsection{Information measures for individual sequences}

\subsubsection{Algorithmic Information (``Algorithmic Complexity")}

     The Shannon entropy measures the average information needed to encode a sequence (i.e., a 
message), but it does not take into account the information needed to specify the encoding procedure 
itself which depends on the probability distribution. This is justified, e.g., if one sends 
repeatedly messages with the same distribution, so that the work involved in setting up the frame 
can be neglected. But the Shannon information tends to underestimate the amount of information 
needed to encode any individual single sequence. On the other hand, in a truly random sample 
there will always be sequences which by chance have a very simple structure, and these sequences 
might be much easier to encode. Finally, not for all sequences it makes sense to consider them as 
members of stationary ensembles. We have already mentioned the digits of $\pi$ and the DNA sequence 
of some specific individual.
      
    To define the information content of individual sequences, we use the fact that any universal 
computer U can simulate any other with a finite emulation program. Thus, if a finite sequence 
$S = s_1 s_2\ldots s_N$ can be computed and printed with a program $Prog_U (S)$ on computer $U$, 
it can be computed on any other computer $V$ by a program $Prog_V (S)$ of length
\be
   L_V(S) \equiv {\rm Len}[Prog_V(S)] \leq {\rm Len}[Prog_U(S)] + c_V \;,     \label{emul}
\ee
where $c_V$ is the length of the emulation program. It is a constant independent of $S$. The 
algorithmic information of $S$ relative to $U$ is defined as the length of the shortest program 
which yields $S$ on $U$,
\be
   C_U (S) = \min_{Prog_U(S)} {\rm Len}[Prog_U(S)] .                        \label{cu}
\ee
If $S$ is infinite, then we can define the algorithmic information per symbol as
\be
                   c(S) = \limsup{N\to\infty} \frac{1}{N} C_U(S_N) \;.
\ee
Notice that $c(S)$ is independent of the computer $U$ due to Eq.~(\ref{emul}), in contrast to the 
algorithmic information of a finite object.
 
    There are some details which we have to mention in connection with Eq.~(\ref{cu}). First of all, 
we have to specify the computer $U$ first. Otherwise, if we would allow to take the minimum in 
Eq.~(\ref{cu}) over all possible computers, the definition would trivially give $C(S) = 1$ for all 
$S$. The reason is that we can always build a computer which on the input ``0" gives S, and for which
all other programs start ``1... " . 

Secondly, we demand that the computer stops after having produced $S$. The reason is essentially 
twofold: On the one hand, we can then define mutual informations very much as in the Shannon
case as $R_c(S,T) = C(S) + C(T) - C(ST)$, where $ST$ is just the concatenation of $S$ and $T$. 
For two uncorrelated sequences $S$ and $T$, this gives exactly zero only if the computer stops 
after having produced $S$. The other reason is that with this convention, we can attribute an 
{\it algorithmic probability} to $S$ by
\be
    P_U(S) = \sum_{Prog_U(S)} 2^{-{\rm Len}[Prog_U(S)]} \;.                   \label{ap}
\ee
In this way, we can define a posteriori a probability measure also in cases where no plausible a 
priori measure exists.

    How is the algorithmic information $C(S)$ related to the Shannon information $h$? In principle, 
we could use the algorithmic probability Eq.~(\ref{ap}) in the definition of h, but we assume instead 
that $S$ is drawn from some other stationary probability distribution, so that $h$ can be defined 
via Eqs.~(\ref{Hn}) -- (\ref{hn}). Then one can show [10] that $C(S) \leq h$ for all sequences, with 
the equality holding for nearly all of them.

    One interesting case where algorithmic and statistical information measures seem to disagree is 
the digits 3141592...  of $\pi$. Since there exist very efficient programs for computing $\pi$ (of 
length $\sim \log N$ for $N$ digits), the algorithmic information of this sequence is zero. But 
these digits look completely random by any statistical criterion [15]: Not only do all digits 
appear equally often, but also all digit pairs, all triples, etc. Does this mean that $\pi$ is random?
     
     The question might sound rather silly. Even if the sequence might {\it look} random, $\pi$ is of 
course not just a random number but carries a very special meaning. Technically, an important 
difference between $C(S)$ and any {\it statistical estimate} of the entropy $h(S)$ is that $C(S)$ 
measures the information needed to specify the {\it first} $N$ digits, while $h$ measures the much 
larger average information needed to specify {\it any} $N$ consecutive digits. For the digits of 
$\pi$, these two are very different: It is much easier to get the first 100 digits, say, than the 
digits from 1001 to 1100.
     For symbol sequences generated by self-­organizing systems, this latter difference is absent.
There, the first digits neither have more meaning nor are in any other way singled out from the 
other digits, whence $\langle C(S)\rangle = h$. The same is true for nearly all
sequences drawn from any ensemble, whence $C(S) = h$ nearly always.

     Thus, for the instances we are interested in, algorithmic information is just a measure of 
randomness, and not a measure of complexity in our sense.

\subsubsection{Ziv-­Lempel Information [16]}

Given an infinite sequence with unknown origin, we never can estimate reliably its algorithmic 
information, since we never know whether we indeed have found its shortest description. Said 
more technically, $C(S)$ is not effectively computable. A way out of this problem is to give up 
the requirement that the description is absolutely the shortest one. Instead, we can arbitrarily 
restrict the method of encoding, provided that this restriction is not so drastic that we get 
trivial results in most cases. In the literature, there exist several suggestions of how to 
restrict the encodings of a sequence in order to obtain effectively computable algorithmic 
information measures. The best known and most elegant is due to Ziv and Lempel [16].

  There, the sequence is broken into words $W_1, W_2, \ldots$ such that $W_0 = \varnothing$, and 
$W_{k+1}$ is the shortest new word immediately following $W_k$. For instance, the sequence $S = 
11010100111101001\ldots$ is broken into $(1)(10)(101)(0)(01)(11)(1010)(01\ldots$. In this way, 
each word $W_k$ with $k>0$ is an extension of some $W_j$ with $j<k$ by one single digit 
$s_{\rm last}$. It is encoded simply by the pair $(j,s_{\rm last})$. It is clear that this is 
a good encoding in the sense that $S$ can be uniquely decoded from the code sequence. Both the 
encoder and the decoder built up the same dictionary of words, and thus the decoder can always 
find and add the new word.

Why is it an efficient code? The reason is that for sequences of low apparent entropy there 
are strong repetitions, such that the average length of the code words $W_k$ increases faster, 
and the number of needed pairs $(j, s_{\rm last})$ increases less than for high-entropy 
sequences. More precisely, the average word length increases roughly like [14]
\be
   \langle {\rm Len}[W_k] \rangle \approx \frac{ \log N}{h}     \label{LZ}
\ee
with the length $N$ of the sequence, and the information needed to encode a pair 
$(j, s_{\rm last})$ increases like $\log N$. The total length of the code sequence is thus 
expected to be $M_N \approx (1+\log N) (N h/\log N) = hN(1+O(1/\log N))$.

More precisely, the Ziv-Lempel information (or Ziv-Lempel ``complexity" \footnote{In [16], 
Ziv \& Lempel introduced another quantity which they called the ``complexity" of their
algorithm, but which they later never used, obviously to avoid further confusion between
that complexity and the Ziv-Lempel information.}) of $S$ is defined via the length $M_N$ of
the resulting encoding of $S_N$ as
\be
        c_{ZL} (S) = \lim_{N\to\infty} M_N/N.        \label{CZL}
\ee
It is shown in ref.(16] that $C_{ZL} = h$ for nearly all sequences, provided $h$ is defined. 
Thus, the Ziv-Lempel information agrees again with the Shannon entropy in the cases we are 
interested in. Indeed, $C_{ZL} = h$ holds also for sequences like the digits of $\pi$ since, 
like $h$, the Ziv-Lempel coding is only sensitive to the statistics of (increasingly long) 
blocks of length $\ll N$.

In practice, Ziv-Lempel coding is a very efficient method of data compression [17], and 
methods related to it (based on Eq.~(\ref{CZL})) are among the most efficient ones for estimating 
entropy [14,18] \footnote{This remark is now obsolete. For an overview over recent activities
in text compression, see http://mattmahoney.net/dc/text.html.}.

Just like Eq.~(\ref{h}) for the Shannon entropy, and like the minimal description length 
entering the definition of the algorithmic information , Eq.~(\ref{CZL}) converges from above. 
Figure 2 shows qualitatively the behavior of the block entropies $H_n$ versus the block 
length $n$, and the Ziv-Lempel code length versus the string length $N$. Both curves have 
the same qualitative appearance, though the interpretation is slightly different. In the 
Shannon case, it is the correlations between symbols which are more and more taken into 
account, so that the information per symbol -- the slope of the graph -- decreases with $n$. 
In the Ziv-Lempel case, on the other hand, in addition to the specific sequence also the 
information about the probability distribution has to be encoded. This contributes mostly at 
the beginning, whence the information per symbol is highest at the beginning. Otherwise stated, 
Ziv-Lempel coding is self-learning, and its efficiency increases with the length of the 
training material.

As we have already mentioned, the Ziv-Lempel encoding is just one example of a wide class of 
codes which do not need any a priori knowledge of the statistics to be efficient. A similar 
method which is more efficient for finite sequences but which is much harder to implement was 
given by Lempel and Ziv [19]. For a recent treatment, see ref. [20].

The increase of the average code length of Ziv-Lempel and similar codes with $N$ has been studied 
for probabilistic models in ref.[20]. In particular, it was shown there that for simple models 
such as moving average or autoregressive models which depend on $k$ real parameters one has
\be
    \langle M_N \rangle \approx   hN + \frac{k}{2} \log N \;.      \label{Riss}
\ee
This is easily understood: For an optimal coding, we have somehow also to encode the $k$ parameters, 
but for finite $N$ we will only need them with finite precision. If the central limit theorem holds, 
their tolerated error will decrease as $N^{-1/2}$, whence we need $\frac{1}{2} \log N$ bits per 
parameter. In this context, we might mention that in the case of algorithmic information, 
including the information about the sequence length (i.e., the information when to stop the
construction) gives a contribution $\sim \log N$ to $C(S_N)$ also for
trivial sequences. This is not yet included in Eq.~(\ref{Riss}).

\begin{figure}

\includegraphics[width=0.43\textwidth]{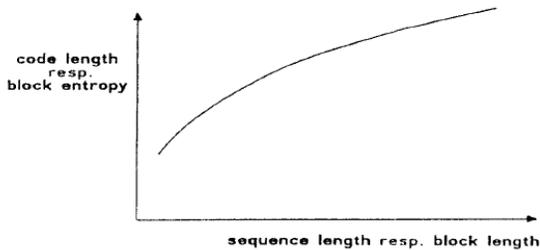}
\caption{Ziv-Lempel code length for a typical sequence with finite entropy versus the sequence 
   length $N$, resp. Shannon block entropies versus block length $n$ (schematically).}

\end{figure}

\section{MEASURES OF COMPLEXITY}

\subsection{General}

   Phenomena which in the physics literature are considered as complex are, among others, chaotic 
dynamical systems, fractals, spin glasses, neural networks, quasicrystals, and cellular automata 
(CA). Common features of these and other examples are the following:

(1) They are somehow situated between disorder and (simple) order [22], i.e. they involve some 
hard to describe and not just random structures. As an example, consider fig.3. There, virtually 
nobody would call the left panel complex. Some people hesitate between the middle and right panels 
when being asked to point out the most complex one. But once told that the right one is created by 
means of a random number generator, the right panel is usually no longer considered as complex -- 
at least until it is realized that a ``random" number generator does not produce random numbers at 
all.

(2) They often involve {\it hierarchies} (e.g. fractals and spin glasses). Indeed, hierarchies 
have often been considered as a main source of complexity (see, e.g., [22]).

(3) But as the example of human societies shows most clearly, a strict hierarchy can be 
ridiculously simple when compared to what is called a ``tangled hierarchy" in ref.~[23]. This 
is a hierarchy violated by feedback from lower levels, creating in this way
``strange loops". Feedback as a source of complexity is also obvious in dynamical systems. 
On the logical (instead of physical) level, in the form of self reference, feedback is the 
basis of Goedel's theorem which seems closely tied to complex behavior [23,10].

\begin{figure}
\includegraphics[width=0.5\textwidth]{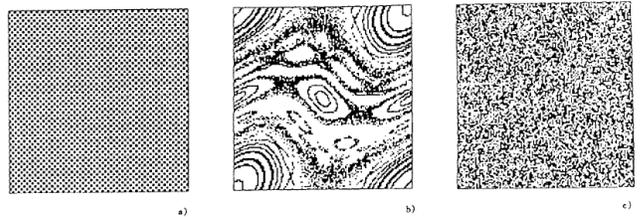}
\caption{ Three patterns used to demonstrate that the pattern that one would intuitively call 
   the most complex is neither the one with the lowest entropy (left) nor the one with the highest 
   (right). That is, complexity is {\it not} equivalent to randomness, but rather is between order
   and chaos.}
\end{figure}

(4) In a particular combination of structure and hierarchy, an efficient and meaningful 
description of complex systems usually requires concepts of different levels. The essence of 
self-generated complexity seems to be that higher-level concepts arise without being put in 
explicitly. 

As a simple example, consider figs.4 to 6. These figures show patterns created by the 
1-dimensional CA with rule nr. 110 in Wolfram's [24] notation, in decreasing resolution. 
In this CA, a row of ``spins" with $s \in \{0,1\}$ is simultaneously updated by the rule 
that neighborhoods 111, 100, and 000 give a ``0", while the other 5 neighborhoods give ``1". 
Figure 4 shows that this CA has a periodic invariant state with spatial period 14, i.e. a 
14-fold degenerate ``ground state". In a first shift from a low-level to higher-level 
description, we might call these ``vacua", although the original vacuum is of course the state 
with zeroes only. Figure 5 shows that between different vacua there are kinks which on a coarser 
scale propagate like particles. In Fig.~6, finally, only the ``particles" are shown, and we see 
a rather complicated evolution of a gas of these particles in which the original concept (the 
spins) are completely hidden. Notice that nothing in the original formulation of the rule had 
hinted at the higher level concepts (vacua and particles).

(5) Complex systems are usually composed of many parts, but this alone does not yet qualify 
them as complex: An ideal gas is not more complex than a human brain because it has more 
molecules than the brain has nerve cells. What is important is that there are strong and 
non-trivial {\it correlations} between these parts. Technically, this is best expressed via mutual 
informations, either probabilistically [26] or algorithmically [21]: A system is complex if 
mutual informations are strong and slowly decaying.

\begin{figure}[htb]
\includegraphics[width=0.45\textwidth,angle=1]{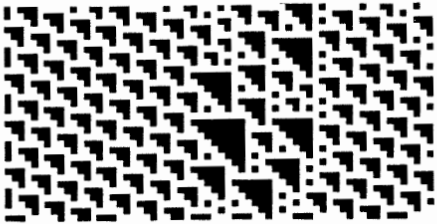}
\caption{Pattern created with CA nr. 110 from a random start. The periodic pattern with spatial 
   periodicity 14 and time periodicity 7 seems to be attractive, but between two phase-shifted 
   such patterns there must be stable kinks. Time increases downward.}
\end{figure}

\begin{figure}
\includegraphics[width=0.5\textwidth,angle=-1]{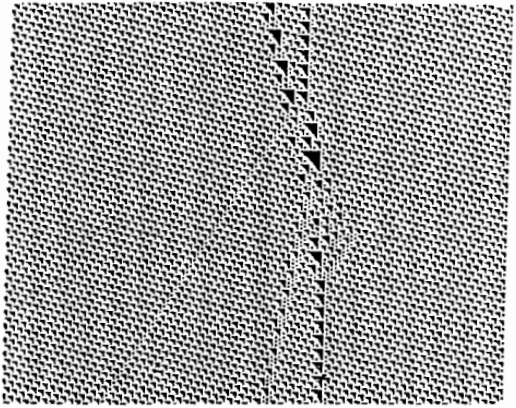}
\caption{Collisions between two kinks for CA 110 [25]. Notice that the kinks behave like 
   ``particles". There are such ``particles" with at least 6 different velocities.}
\end{figure}

\begin{figure}
\includegraphics[width=0.5\textwidth,angle=0]{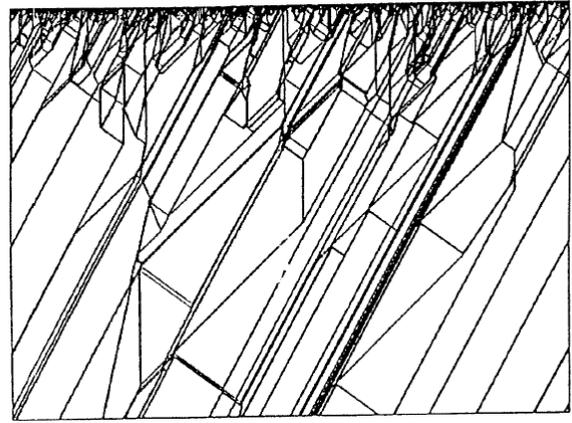}
\caption{Evolution of CA 110 on a coarse-grained scale. Every pixel represents a block of 14 
   spins. The pixel is white if the block is the same as the neighboring block to the left,
   otherwise it is black. In this way, only kinks are seen. In order to compress vertically, only
   every 21-th time step is shown.}
\end{figure}

Examples are of course fractals and critical phenomena, but the correlations there seem still 
very simple compared to those in some cellular automata. 

First, there are computationally universal automata like the ``game of life". For them, we can 
construct initial configurations which do whatever function we want them to do. But these 
configurations are in general rather tricky, with strong constraints over very long distances. 
In Fig.~7, we show a configuration which is called a ``glider gun" [27]. Periodically, it sends 
out a ``glider" which then moves -- oscillating with period 4 -- diagonally away from the gun. 
In this configuration, all black cells in the upper part are needed in order to function, i.e. 
there are very strong initial correlations if the state is to do what we want it to do.

\begin{figure}
\includegraphics[width=0.5\textwidth]{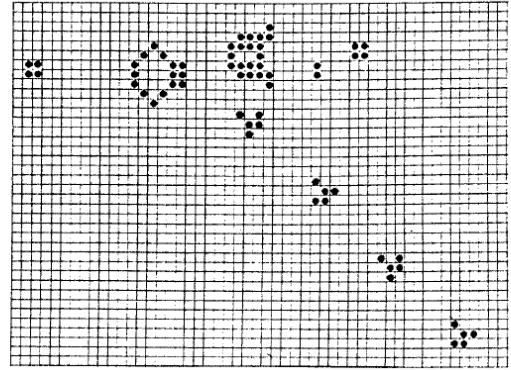}
\caption{A ``glider gun" in the game of life. The game of life is defined so that in 
   each generation, a cell becomes black (``alive") if it had in its nine-cell neighborhood 
   3 or 4 alive cells in the previous generation. Else it dies. Gliders are the configurations 
   made up of 5 alive cells in the lower right corner (from ref.[27]).}
\end{figure}

While Fig.~7 illustrates the correlations in configurations specially designed to do something 
meaningful, in another cellular automaton strong correlations seem to appear spontaneously. 
This is the simple 1-dimensional CA with rule 22 [28]. In this model, which will be treated in 
more detail in the last section, any random initial configuration leads to a statistically 
stationary state with extremely long and hard to describe correlations. The non-triviality of 
these correlations is best expressed by the fact that they neither seem to be self-similar or 
fractal, nor quasi-periodic, nor ordered (block entropies $H_n$ tend to infinity for increasing 
block lengths), nor random (entropy estimates $h_n = H_{n+1} - H_n$ seem to tend towards zero).

The correlations between the parts of a complex object imply that the whole object in a way 
is ``more than its parts". Let me be more specific. Of course, we can describe the whole object 
completely by describing its parts, and thus it cannot be more than these. But such a description 
will be redundant, and it misses the fact that the correlations probably indicate that the 
object as a whole might have some meaning.

(6) More important than correlations among the parts of a complex system are often correlations 
between the object and its environment [29,30].

Indeed, a piece of DNA is complex not so much because of the correlations within its base pairs, 
but because it matches with a whole machinery made for reading them, and this machinery 
matches -- i.e., is correlated with -- the rest of the organism, which again matches with the 
environment in which the organism is to live in. Similarly, the letters of a novel in some 
language could hardly be considered as complex if there were no possibility in principle to read 
it. The ability to read a novel, and to perceive it thus as complex, requires some correlation 
between the novel and the reader.

This remark opens several new questions. First, it shows that we have to be careful when talking 
about the complexity ``of an object". Here, we obviously would rather mean the complexity increase 
of the combined system \{object + environment\} when adding the object. Secondly, when saying that 
correlations with structures within ourselves might be crucial for us to ``perceive" an object as 
complex, I did not mean this subjectivity as derogatory. Indeed, I claim (and shall comment more 
on this later) that it is impossible to define a purely objective notion of complexity. Thus, 
complexity only exists if it has a chance to be {\it perceived}. Finally, the above shows that 
complexity is related to {\it meaning} [31].

(7) Pursuing seriously the problem that complexity is somehow related to the concept of {\it meaning} 
would lead us too far from physics into deep questions of philosophy, so I will not do it. But 
let me just point out that a situation acquires some meaning to us if we realize that only some 
of its features are essential, that these features are related to something we have already stored 
in our memory, or that its parts fit together in some unique way. Thus we realize that we can 
replace a full, detailed and straightforward description by a compressed and more ``intelligent" 
description which captures only these essential aspects, eventually employing already existing 
information. We mention this since the relation between compression of information and complexity 
will be a recurrent theme in the following. {\it Understanding} the meaning is just the act of 
replacement of a detailed description containing all irrelevant and redundant information by a 
compressed description of the relevant aspects only.

(8) From the above we conclude that complexity in some very broad sense always is a {\it difficulty 
of a meaningful task}. More precisely, the complexity of a pattern, a machine, an algorithm, etc., 
is the difficulty of the most important task related to it.

By ``meaningful" we exclude on the one hand purely mechanical tasks, as for instance the lifting 
of a heavy stone. We do not want to relate this to any complexity. But we also want to exclude the 
difficulty of coding, storing, and reproducing a pattern like the right panel of fig.3, as the 
details of that pattern have no meaning to us.

A technical problem is that, when we speak about a difficulty, we have to say what are our allowed 
tools and what are our most important limitations. Think e.g. of the complexity of an algorithm. 
Depending on whether CPU time is most seriously limited or core memory, we consider the time or 
space complexity as the more important. Also, these complexities depend on whether we use a 
single-CPU (von Neumann) or multiple-CPU computer.

(9) Another problem is that we don't have a good definition of ``meaning", whence we cannot take 
the above as a real definition of complexity. This is a problem in particular when we deal with 
self-generated complexity.

In situations which are controlled by a strict outside hierarchy, we can replace ``meaningful" by 
``functional", as complex systems are usually {\it able to perform some task} [31]. This is also true 
in the highest self-organized systems, namely in real life. There it is obvious, that e.g. the 
complexity of an eye is due to the fact that the eye has to perform the task of seeing.

  But in 
systems with self-generated complexity it is not always obvious what we really mean by a ``task". 
In particular, while we often can see a task played by some part of a living being (the eye) or 
even of an ecosystem, it is impossible to attribute a task to the entire system. Also, we must 
be careful to distinguish between the mere ability to perform a task, and the
{\it tendency} or {\it probability} to do so.

Let me illustrate this again with cellular automata. As we said, the ``game of life" can be 
programmed as a universal computer. This means that it can simulate any possible behavior, given 
only the right initial configuration. For instance, we can use it to compute the digits of $\pi$, 
but we can also use it to proof Fermat's last theorem (provided it is right). Thus it can do 
meaningful tasks if told so from the outside. But this does not mean that it will do interesting 
things when given a random initial condition. Indeed, for most initial conditions the evolution 
is rather disappointing: For the first 400 time steps or so, the complexity of the pattern seems 
to increase, giving rise (among others) to many gliders, but then the gliders collide with all 
the interesting structures and destroy them, and after about 2000 time steps only very dull 
structures survive. This is in contrast to real life (and, for that part, to CA rule 22!), which 
certainly has also started with a random initial configuration, but which is still becoming more 
and more complex.

(10) As a consequence of our insistence on meaningful tasks, the concept of complexity becomes 
{\it subjective}. We really cannot speak of the complexity of a pattern without reference to the observer. 
After all, the right-hand pattern in fig.3 might have some meaning to somebody, it is just we who 
decide that it is meaningless.

This situation is of course not new in physics. It arises also in the Copenhagen interpretation 
of quantum mechanics, and it appears also in Gibbs' paradoxon. In the latter, the entropy of an 
isotope mixture depends on whether one wants to distinguish between the isotopes or not. Yet it 
might be unpleasant to many, in particular since the dependence on the observer as regards complexity 
is much less trivial than in Gibbs' paradoxon.

Indeed, statistical mechanics suggests an alternative to considering definite object as we 
pretended above. It is related to the observation that when we call the right panel of fig.3 
random, we actually do not pay attention to the fine details of that pattern. We thus do not 
really make a statement about that very pattern, but about the class of all similar patterns. 
More precisely, instead saying that the pattern is not complex, we should (or could, at least) 
say ``that pattern seems to belong to the class of random patterns, and this class is trivial 
to characterize: it has no features" (32). The question of what feature is ``meaningful" is now 
replaced by the question of what ensemble to use.

We have thus a dichotomy: We can either pretend to deal with definite objects, or we can pretend 
to deal only with equivalence classes of objects or probability distributions (``ensembles"). In 
the latter case we avoid the problems of what is ``meaning" by simply defining the ensembles we 
want to study. This simplifies things somewhat at the expense of never knowing definitely whether 
the objects we are dealing with really belong to the ensemble, resp. whether they aren't objects 
appearing with zero probability. This dichotomy corresponds exactly to the two ways of defining 
information, discussed in the previous section.

In the following, the dichotomy will be seen more precisely in several instances. In the 
tradition of physics, I will usually prefer the latter (ensemble) attitude. Also in the tradition 
of physics and contrary to the main attitude of computer science, I will always stress 
probabilistic aspects in contrast to purely algorithmic ones. Notice that the correlations 
mentioned under point (5) are defined most easily if one deals with probability distributions. 
For a conjecture that correlations defined not probabilistically but purely algorithmically are 
basic to a mathematical approach to life, see ref. [21].

In this way, the complexity of an object becomes a {\it difficulty related to classifying the object, 
and to describing the set or rather the ensemble to which it belongs}. 

These general considerations have hopefully sorted out somewhat our ideas what a precise 
definition of complexity should be like. They have made it rather clear that a unique definition 
with a universal range of application does not exist (indeed, one of the most obvious properties 
of a complex object is that there is no unique most important task associated with it). Let us 
now take the above as a kind of shopping list, let us go to the literature, and let us see how the 
things we find there go together with it.

\subsection{Space and Time Complexity of Algorithms [33]}

We have already mentioned shortly the complexity of algorithms. The space and time complexities of 
an algorithm are just the required storage and CPU time, respectively. A family of problems depending on a 
discrete parameter $N$ is considered complex (more precisely ``NP-hard") if the fastest algorithm 
solving the problem for every $N$ needs a time which increases exponentially, although the 
formulation and verification of a proposed solution would at most increase polynomially.

Although this is of course a very useful concept in computer science, and moreover fits perfectly 
into our broad definition of complexity, it seems of no relevance to our problem of self-generated 
complexity. The reason is that an algorithm is never self-generated but serves a purpose imposed 
from outside. 

Thus, the computational complexity of an algorithm performed {\it by} a supposedly complex 
system (e.g., by a bat when evaluating echo patterns) cannot be used as a measure for the complexity 
of the system (of the bat). However, most of the subsequent complexity measures are related to 
complexities of algorithms {\it we} are supposed to perform if we want to deal with the system.

\subsection{Algorithmic and Ziv-Lempel Complexities}

It is clear that these information measures are also complexity measures in the broad sense defined 
above: they measure a difficulty of a task, namely the task of storing and transmitting the full 
description of an object. They differ just in the tools we are allowed to use when describing 
(``encoding") the object.

The problem why we hesitate to accept them as relevant complexity measures is that in the cases 
we are interested in, a detailed description of the object is not meaningful. Take for instance a 
drop of some liquid. Its algorithmic complexity would be the length of the shortest description of 
the positions of all atoms with infinite precision -- a truly meaningless task! Indeed, on an even 
lower level, we would have to describe also all the quantum fluctuations, and the task becomes 
obviously impossible. But the algorithmic complexity of this drop is completely different if we 
agree to work not with individual drops but with ensembles; the task to describe the drop in the 
canonical ensemble, say, is drastically reduced. And it is again different if we are content with 
the grand canonical ensemble.

The situation is less dramatic but still very similar for symbol sequences generated by dynamical 
systems. On the formal level, this is suggested by the applicability of the thermodynamic 
formalism [34] to dynamical systems. On the more important intuitive level, this is best seen from 
the example of the weather considered as a dynamical system. There, certainly the task of storing 
and transmitting descriptions of climatic time sequences is much less relevant than e.g. the task 
of forecasting, i.e. of finding and using the correlations in these sequences.

In other cases like the digits of $\pi$ or the letters of a well-written computer program, we would 
accept the algorithmic information is a perfectly reasonable complexity measure. These sequences are 
meaningful by themselves, and when asking about their complexities we probably do not want to take 
them just as representatives of some statistical ensembles.

This illustrates indeed best the subjectivity of any complexity measure mentioned above. The fact 
that complexity is ``just" subjective was very often realized, but usually this is not considered as 
deep. Instead, it is considered in general as a sign that ``the correct" definition of complexity is 
not yet found. I disagree. It is common to all science that it does not deal with reality but with 
idealizations of it. Think just of free fall, which -- though never observed exactly on Earth -- is 
an extremely useful concept. In the same way, both concepts -- of an object by itself being 
meaningful, and representing only an ensemble -- are useful idealizations.

Very often, different details of a system are separated by large length scales. In these cases, 
it seems obvious what one wants to include in the description, and the subjectivity of the
complexity is less of a problem. It is more important in those cases which we consider naively 
as most complex: There either the length scale separations are absent, or they are messed up by 
feedback or by amplification mechanisms as in living beings.

\subsection{Logical Depth [35]}

A complexity measure more in our spirit is the ``logical depth" introduced by C. Bennett [35]. 
The logical depth of a string $S$ is essentially the time needed for a general purpose computer 
to actually {\it run the shortest program} generating it. Thus the task is now not that of 
storing and retrieving the shortest encoding, but that of actually performing the decoding. 
The difference with time complexity (Sec.3b) is that now we do not ask for the time needed by 
the {\it fastest} program, but rather the {\it shortest}.

The reason why this is a good measure, in particular of self- generated complexity, is Occam's 
razor: If we find a complex pattern of unknown origin, it is reasonable to assume that it was 
generated from essentially the shortest possible program. The program must have been assembled by 
chance, and the chance for a meaningful program to assemble out of scratch decreases exponentially 
with its length.

For a random string $S$, the time needed to generate it is essentially the time needed to read in 
the specification, and thus it is proportional to its length. In contrast to this, a string with 
great logical depth might require only a very short program, while decoding the program takes very 
long, much longer than the length of $S$. The prime example of a pattern with great logical depth 
is presumably life [35]. As far as we know, life emerged spontaneously, i.e. with a ``program" 
assembled randomly which had thus to be very short. But it has taken some $10^9$ years to work with 
this program, on a huge parallel analog computer called ``earth", until life has assumed its 
present forms.

A problem in the last example is that ``life" is not a single pattern but rather an ensemble. Noise 
from outside was obviously very important for its evolution, and it is not at all clear whether we 
should include some of this noise as ``program" or not. The most reasonable attitude seems to be 
that we regard everything as noise which from our point of view does not seem meaningful, and 
consider as program the rest. Take for instance the environmental fluctuations which lead to the 
extinction of the dinosaurs. For an outside observer, this was just a large noise fluctuation. 
For us, however, it was a crucial event without which we would not exist. Thus we must consider it 
part of our ``program".

A more formal example with (presumably) large logical depth is the central vertical column in Fig. 8. 
This figure was obtained with cellular automaton nr. 86, with an initial configuration consisting 
of a single ``1". Since both the initial configuration and the rule are very easy to describe, the 
central column has zero Kolmogorov complexity. From very long simulations it seems however that it 
has maximal entropy [36]. Furthermore, it is believed that there exists no other way of getting 
this column than by direct simulation. Since it takes $\propto N^2$ operations to iterate $N$ time 
steps, we find indeed a large logical depth.

\begin{figure}
\includegraphics[width=0.5\textwidth,angle=0]{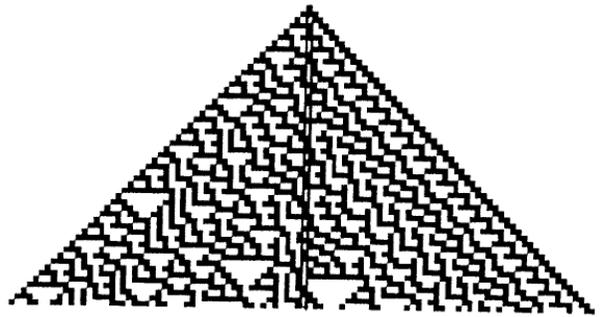}
\caption{Pattern generated by CA \# 86, from an initial configuration having a single ``1". The 
central column (marked by 2 vertical lines) seems to be a logically deep sequence.}
\end{figure}

Formally, Bennett defines a string to be $d$-deep with $b$ bits of significance, if every program 
to compute it in time $\leq d$ could be replaced by another program which is shorter by $b$ bits. 
Large values of $d$ mean that the most efficient program -- from the point of view of program 
length -- takes very long to run. The value of $b$ is a significance very similar to the 
statistical significance in parameter estimation: If $b$ is small, then already a small change 
in the sequence or in the computer used could change the depth drastically, while for large $b$ 
it is more robust.

A more compact single quantity defined by Bennett is the {\it reciprocal mean reciprocal depth}
\be
   d_{\rm rmr}(S) = P_U(S) \left[ \sum_{Prog_U(S)} \frac{2^{-{\rm Len}(Prog_U(S))}}{t(S)}\right]^{-1}\;,
\ee
where $P_U(S)$ is the algorithmic probability of $S$ defined in Eq.~(\ref{ap}), and $t(S)$ is the 
running time of the program $Prog_U(S)$. The somewhat cumbersome use of 
reciprocals is necessary since the alternative sum with $1/t(S)$ replaced by $t(S)$ would be 
dominated by the many slow and large programs.

Logical depth shares with algorithmic information the problem of machine-dependence for finite 
sequences, and of not being effectively computable [35]. We never know whether there isn't a 
shorter program which encodes for some situation, since this program might take arbitrarily long 
to run. What we use here is the famous undecidability of the halting problem. We can however -- 
at least in principle -- exclude that any shorter problem needs less time to run. Thus, we can 
get only upper bounds for the algorithmic information, and lower bounds on the logical depth.

\subsection{Sophistication [37]}

We have already seen in Sec. 2b (see Fig.~2) that description length increases sublinearly with 
system size (is a convex function). This can also be seen in the following way.

In practical computers there is a distinction between program and data. While the program specifies 
only the class of patterns, the data specify the actual object in the class. A similar distinction 
exists in data transmission by conventional (e.g., Huffman [38]) codes, where one has to transmit 
independently the rules of the code and the coding sequence.

It was an important observation by Turing that this distinction between program and data is not 
fundamental. The mixing of both is e.g. seen in the Ziv-Lempel code, where the coding sequence and 
the grammatical rules used in compressing the sequence are not separated. For a general discussion 
showing that the rule {\it vs.} data separation is not needed in communication theory, see ref. [20].
The convexity mentioned above is due to this combination of ``data" and ``program". If the 
``program" and the algorithmic information are both finite, we expect indeed that the combined code
length $M_N$ increases asymptotically like $M_N = const + hN$, where the additive constant is the 
program length. This is shown schematically in Fig.~9. The offset of the asymptotically tangent 
line on the y-axis is the proper program length.

It was shown by Koppel and Atlan [37] that this is essentially correct. The length of the proper 
program, called by them ``sophistication", can moreover be defined such that it is indeed independent 
of the computer $U$ used in defining $M_N$.

Sophistication is a measure of the importance of rules in the sequence. It is a measure of the 
difficulty to state those properties which are not specific to individual sequences but to the entire 
ensemble. Equivalently, it is a measure of the importance of correlations. Rules imply correlations, 
and correlations between successive parts of the sequence $S$ imply that the description of a previous 
part of $S$ can be re-used later, reducing thus the overall program length. This aspect of complexity 
in an algorithmic setting had been stressed before in ref.[21].

\begin{figure}
\includegraphics[width=0.5\textwidth,angle=0]{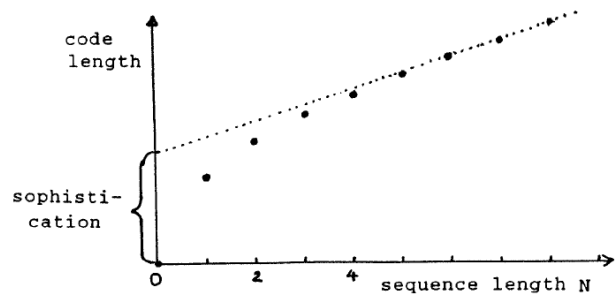}
\caption{Total program length for a typical sequence with finite algorithmic information per digit, 
   and with finite proper program length (``sophistication"; schematically).}
\end{figure}

As we said already, the increase of average code length with $N$ has been studied for probabilistic 
models in [20], with the result that the leading contributions are given by Eq.~(\ref{Riss}). While 
this equation and its interpretation fit perfectly into our present discussion, it shows also a 
weakness which sophistication shares with most other complexity measures proposed so far. It shows 
that unfortunately sophistication is infinite (diverging $\sim \log N$) already in rather simple 
situations. More precisely, it takes an infinite amount of work to describe any typical real number. 
Thus, if the system depends on the precise value of a real parameter, most proposed complexity 
measures must be infinite.

\subsection{Effective Measure Complexity [32]}

Let us now discuss a quantity similar in spirit to sophistication, but formulated entirely within 
Shannon theory. There, one does distinguish between rules and data, with the idea that the rules are 
encoded and transmitted only once, while data are encoded and transmitted again and again. Thus the 
effort in encoding the rules is neglected.

The average length of the code for a sequence of length $n$ is the block entropy $H_n$ defined in 
Eq.~(\ref{Hn}). We have already pointed out that they are also convex like the code lengths $M_n$ in 
an algorithmic setting, and thus thus their differences
\be
   h_n      = H_{n+1}    - H_n
\ee
are monotonically {\it de-}creasing to the entropy $h$. When plotting $H_n$ versus $n$ (see Fig.~2), 
the quantity corresponding now to the sophistication is called {\it effective measure complexity} (EMC)
in [32] \footnote{It was indeed first introduced by R. Shaw in {\it The dripping faucet as a model 
chaotic system}, Aerial Press, Santa Cruz 1984, who called it ``excess entropy". Claims made by 
Crutchfield {\it et al.} [Phys. Rev. Lett. {\bf 63}, 105 (1989); Physica D {\bf 75}, 11 (1994);
CHAOS {\bf 13}, 25 (2003)] that it was first introduced in J. P. Crutchfield and N. Packard,
Physica D {\bf 7}, 201 (1983) are wrong.}
\bea
   EMC & = & \lim_{n\to\infty} [ H_n- n(H_n-H_{n-1})]   \nonumber \\
       & = & \sum_{k=0}^\infty (h_k - h) \;.                     \label{3.3}
\eea
The EMC has a number of interesting properties. First of all, within all stochastic processes with 
the same block entropies up to some given $n$, it is minimal for the Markov process of order $n-1$ 
compatible with these $H_k$. This is in agreement with the the notion that a Markov ansatz is the 
simplest choice under the constraint of fixed block probabilities. In addition, it is finite for 
Markov processes even if these depend on real parameters, in contrast to sophistication. It has 
thus the tendency to give non-trivial numbers where other measures fail.

Secondly, it can be written as a sum over the non-negative decrements $\delta h_n = h_{n-l}-h_n$ as
\be
   EMC = \sum_{n=1}^\infty n\delta h_n.
\ee
The decrement $\delta h_n$ is just the average amount of information by which the uncertainty of 
$s_{n+1}$ decreases when learning $s_1$, and when all symbols $s_k$ between are already known. 
Thus EMC is the average usable part of the information about the past which has to be remembered 
at any time if one wants to be able to reconstruct the sequence $S$ from its shortest encoding, 
which is just the mutual information between the past and future of a bi-infinite string. 
Consequently, it is a lower bound on the average amount of information to be kept about the past, 
if one wants to make an optimal forecasting. The latter is obviously a measure of the {\it difficulty 
of making a forecast}.

Finally, in contrast to all previous complexity measures it is an effectively computable observable, 
to the extent that the block probabilities $p_N(s_1\ldots s_N)$ can be measured.

The main drawback of the EMC in comparison to an algorithmic quantity like sophistication is of 
course that we can apply it only to sequences with stationary probability distribution. This 
includes many interesting cases, but it excludes many others.

\subsection{Complexities of Grammars [33]}

A set of sequences (or ``strings") over a finite ``alphabet" is usually called a formal language, 
and the set of rules defining this set is called a ``grammar". In agreement with our general 
remark that complexities are preferably to be associated to sets or ensembles, it is natural to 
define a complexity of a grammar as the difficulty to state and/or apply its rules.

There exists a well-known hierarchy of formal languages, the Chomsky hierarchy [33]. Its main 
levels are in increasing complexity and generality: {\it regular languages, context-free languages, 
context-sensitive languages}, and {\it recursively enumerable sets}. They are distinguished by the 
generality of the rules allowed in forming the strings, and by the difficulty involved in testing 
whether some given string belongs to the language, i.e. is ``grammatically" correct.

Regular languages are by definition such that the correctness can be checked by means of a finite 
directed graph. In this graph, each link is labeled by a symbol from the alphabet, and each symbol 
appears at most once on all links leaving any single node (such graphs are called ``deterministic"; 
any grammar with a finite non-deterministic graph can be replaced by a deterministic graph). 
Furthermore, the graph has a unique start node and a set of stop nodes. Any grammatically correct 
string is then represented by a unique walk on the graph starting on the start node and stopping on 
one of the stop nodes, while any wrong string is not. For open-ended strings, we declare each node 
as a stop node, so that all strings are accepted which correspond to allowed walks. Scanning the 
string consists in following the walk on the graph.

\begin{figure}
\includegraphics[width=0.5\textwidth,angle=0]{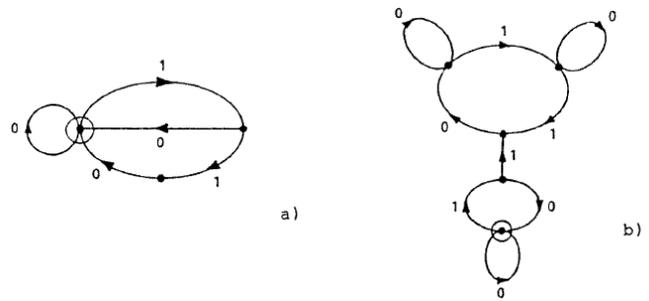}
\caption{Deterministic graphs for the regular languages generated in 1 time step by CA rules nr. 
   76 (a) and 18 (b). The heavy nodes are the start nodes (From ref. [39]).}
\end{figure}

Examples of graphs for regular languages are given in Fig.~10. They correspond to strings allowed 
in the second generation of two cellular automata, if any string is allowed as input in the first 
generation. Figure 10(a) corresponds e.g. to the set of all strings without blocks of three 
consecutive ``1"s, and with no further restriction.

One might define the complexity of the grammar as the difficulty to write down the rules, i.e. 
essentially the number of nodes plus the number of links. However, in ref. [39] the regular 
language complexity (RLC) was defined as
\be
    RLC = \log n , 
\ee
where $n$ is the number of nodes alone, of the smallest graph giving the correct grammar (usually, 
the graph of a grammar is not unique [33]). This makes indeed sense as the so defined RLC is 
essentially the difficulty in performing the scan: during a scan, one has to remember the index of 
the present node, in order to look up the next node(s) in a table, and then to fetch the index of 
the next node. If no probabilities are given, the average information needed to fetch a number 
between 1 and $n$ (and the average time to retrieve it) is $\log n$.

Assume now that one is given not only a grammar but also a stationary probability distribution, 
i.e. a stationary {\it ensemble}. This will also induce probabilities $P_k (k=1,...n)$ for being at 
the $k$-th node of the graph at any given time. Unless one has equidistribution, this will help in 
the scan. Now, both the average information about the present node and the time to fetch the next 
one will be equal to the ``{\it set complexity}" (SC),
\be
   SC = - \sum_{k=1}^n P_k \log P_k .       \label{SC}
\ee
It is obvious that the SC is never larger than the RLC, and is finite for all regular languages. 
It is less obvious that while the RLC is by definition infinite for all other classes in the 
Chomsky hierarchy, the same need not hold for the SC. All context-free and context-sensitive 
languages can be represented by infinite non-deterministic graphs [40]. And very often one finds 
deterministic graphs with measures such that the $P_k$ decrease so fast with the distance from 
the start node that the SC is finite [32].

\begin{figure}
\includegraphics[width=0.5\textwidth,angle=0]{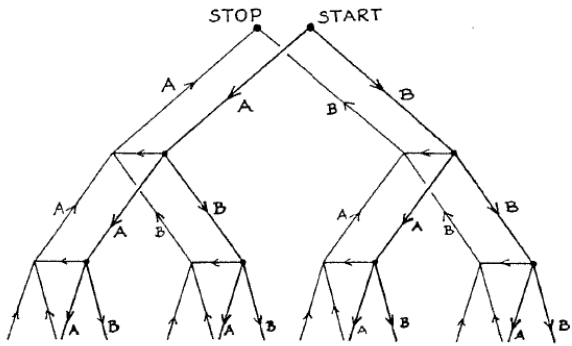}
\caption{Part of a non-deterministic infinite graph accepting all palindromes made up of 2 
   letters A and B. The front root node is the start node, the rear is the stop node.}
\end{figure}

We illustrate this for the set of {\it palindromes}. A palindrome is a string which reads forward
and backward the same, like the letters Adam said when he first met Eve: ``MADAM I'M ADAM" (by the 
way, the answer was also a palindrome: ``EVE"). A non-deterministic graph accepting all palindromes 
built of two letters A,B is shown in Fig.~11. It has a tree-like structure, with the start and 
stop nodes at the root, and with each pair of vertices connected by {\it two} directed links: One 
pointing towards the root, the other pointing away from it. The tree is infinite in order to allow 
infinitely long words (palindromes form a non-regular context-free language; it is shown in [40] 
that for all context-free languages one has similar tree-like graphs). But if the palindromes are 
formed at random, long ones will be exponentially suppressed, and the sum in Eq.~(\ref{SC}) will 
converge.

\subsection{Forecasting Complexity [32,41]}

Both the RLC and the SC can be considered as related to a restricted kind of forecasting. Instead 
of just scanning for correctness, we could have as well forecasted what symbol(s) is resp. are 
allowed to appear next. In a purely algorithmic situation where no probabilities are given, this is 
indeed the only kind of meaningful forecasting.

But if one is given an ensemble, it is more natural not only to forecast what symbols {\it might} 
appear next, but also to {\it forecast the probabilities} with which they will appear. We call 
forecasting complexity (FC) the average amount of information about the past which has to be stored 
at any moment, in order to be able to make an optimal forecast.

Notice that while the Shannon entropy measures the {\it possibility} of a good forecast, the FC 
measures the {\it difficulty} involved in doing so. That these need not be correlated is easily seen 
by looking at left-right symbol sequences for quadratic maps (the symbol ``C" appears for nearly all 
start values $x_0$ with zero probability, and can thus be neglected in probabilistic arguments). For 
the map $x' = 2 - x^2$, e.g., {\it all} R-L sequences are possible [7] and all are equally probable. 
Thus, no non-trivial forecasting is possible, but just for that reason the best forecast is very easy: 
It is just a guess. In contrast, at the Feigenbaum point [5] the entropy is zero and thus perfect 
forecasting is possible, but as shown below, the average amount of information about the past needed 
for an optimal forecast is infinite.

Notice that the FC is just the average logical depth per symbol, when the latter is applied to 
infinite strings drawn from a stationary ensemble. Assume we want to reconstruct such an infinite 
string from its shortest code. Apart from an overhead which is negligible in the limit of a long 
string, we are provided only an information of $h$ bits per symbol ($h$ = entropy), and we are 
supposed to get the rest of 1-$h$ bits per symbol from the past (we assume here a binary alphabet). 
But this involves exactly the same difficulty as an optimal forecast.

In addition to be related to the logical depth, the FC is also closely related to the EMC. As discussed 
in Sec. 3f, the EMC is a lower bound on the average amount of information which has to be stored at any 
time, since it is the time-weighted average of the amount of information by which the uncertainty about 
a future symbol decreases when getting to know an earlier one. Thus, we have immediately [32]
\be
    EMC \leq FC        .   
\ee
Unfortunately, this seems to be a very weak inequality in most cases where it was studied [32,41].

The difficulty of forecasting the Feigenbaum sequence mentioned above comes from the fact that 
Eq.~(\ref{3.3}) for the EMC is logarithmically divergent there.

The forecasting complexity is infinite in many seemingly simple cases, e.g. for sequences 
depending on a real parameter: For an optimal forecast, one has at each time to manipulate an 
infinity of digits. In such cases, one has to be less ambitious and allow small errors in the 
forecast. One expects -- and finds in some cases [41,42] -- interesting scaling laws as the errors 
tend towards zero. Unfortunately, there are many ways to allow for errors [41,42], and any 
treatment becomes somewhat unaesthetic.

\subsection{Complexities of Higher-Dimensional Patterns}

Up to now, we have only discussed 1-dimensional strings of symbols. It is true that any definable 
object can be described by a string, and hence in principle this might seem enough. But there are 
no unique and canonical ways to translate e.g. a two-dimensional picture into a string. A scanning 
along horizontal lines as used in TV pictures, e.g., leads to long range correlations in the string 
which do not correspond to any long range correlations in the picture. Thus, the complexity of the 
describing string is in general very different from what we would like to be the complexity of the 
pattern.

One way out of this is to minimize over all possible scanning paths when applying the above 
complexity measures (the entropy of a two-dimensional discrete pattern can for instance be 
inferred correctly from a string obtained via a Peano curve; moving along such a curve, one 
minimizes the spurious long range correlations). In using this strategy for complexity estimates, 
one should allow multiple visits to sites in order to pick up information only when
it is needed [32]. Forecasting complexity, e.g., would then be defined as the infimum over the 
average amount of information to be stored, if one scans somehow through the pattern and forecasts 
at each site the pixel on the basis of the already visited pixels.

I might add that by ``forecasting" we mean the prediction of the next symbol, provided one has 
already understood the grammar and the probability distribution. The latter is in general not easy, 
and one might justly consider the difficulty of {\it understanding} an ensemble of sequences (or of 
any other systems) as the most fundamental task. We do not discuss it here further since it does not 
easily lend itself to formal discussion.

\subsection{Complexities of Hierarchy Trees}

In a series of papers, Huberman and coworkers [22,43-45] have discussed the complexity of rooted 
trees representing hierarchies. As we have mentioned above, we do not necessarily consider strictly 
hierarchical systems as the most complex ones, yet many systems (including, e.g., grammars for 
formal languages) can be represented by trees and it is an interesting question how to measure 
their complexity.

The starting point of Ref. [43] was very much the same as ours: neither ordered nor completely 
random trees should be considered complex. This left the authors with the notion that the complexity 
of a tree is measured by its lack of self-similarity. This was strengthened later by the 
observation that ``ultra-diffusion" is fastest for non-complex trees [44], and that percolation 
is easiest for them [45]. In this, we have to compare trees with the same ``silhouette", i.e. with 
the same average number of branches per generation.

We can indeed interpret the results of refs.[44,45] slightly different and more in the spirit of 
the above approaches: instead of the lack of self-similarity, it is the amount of correlations
between ancestors and descendants which slows down ultradiffusion and hinders percolation. Thus, 
we see again that correlations are a good measure of complexity. 

A certain drawback of this 
approach is that the quantitative measures proposed in refs.[22,43] seem somewhat arbitrary 
(they are not related to the difficulty of any obvious task in a quantitative way), while no 
quantitative measures are used at all in refs.[44,45] at all.

\subsection{Thermodynamic Depth}

In a recent paper [30], Lloyd and Pagels tried to define a physical measure of complexity very much 
in the spirit of Bennett's logical depth, but which should be an effectively computable observable 
in contrast to logical depth.

I must confess that I have some problems in understanding their approach. Like all the other 
authors mentioned above, they want an observable which is small both for completely ordered 
and for completely random systems. They claim that the right (indeed the unique!) measure 
``must be proportional to the Shannon entropy of the set of trajectories that experiment 
determines can lead to the state" (ref.[30], p.190). Thus, a system is called deeper if it 
has more prehistories than a shallow one. By this criterium, a stone (which could be a petrified 
plant) would be deeper than a plant which cannot be a herbified stone!

This problem might be related to the fact that nowhere in ref.[30] is stated how far back 
trajectories have to be taken. In a chaotic system, this is however crucial. If we agree that we 
always go back until we reach the ``original" blueprint, then the above problem could be solved: 
The first blueprint for a plant dates back to the origin of life, while the stone might be its 
own blueprint. But neither seems this to be understood in ref. [30] nor would it be always clear 
what we consider the ``blueprint".

In table 1, we summarize the various proposed complexity measures and the tasks they are related 
to. As a rule, we can say that for each there exist a probabilistic and an algorithmic version. 
While the algorithmic one is always deeper and much more general, it is only the probabilistic 
one which is computable and measurable, and is thus more directly applicable.

\begin{table*}
\caption{Complexity measures discussed in this paper and tasks they are related to}
\begin{tabular}{l|l}
\hline
\hline

   Task                           &     Complexity Measure   \\  \hline

{\bf perform most efficient}      &  time c. (if CPU time is limiting factor)     \\
               algorithm          &  space c. (if memory space is limiting)   \\
\\
{\bf store} and {\bf retrieve} the  &  algorithmic (Kolmogorov-Chaitin) c.\\
shortest code                     &     \\
\\
store and retrieve, but           &  Ziv-Lempel type c.  \\
coding is {\bf restricted}        &     \\
\\
{\bf decode shortest} code;       &   logical depth (Bennett) \\
perform shortest algorithm        &     \\
\\
{\bf describe set} of symbol strings &  sophistication (Koppel-Atlan)\\
\\
{\bf verify} that symbol string   &    regular language complexity \\
belongs to some regular language  &     \\
\\ 
verify as above, but assuming     & set complexity  \\
stationarity and using a          &     \\ 
{\bf known probability measure}   &     \\
\\
{\bf forecast} a symbol string    &     forecasting c.  \\
chosen randomly out of            &     (probabilistic version of logical depth)\\
stationary ensemble;              &     \\
encode its shortest description   &     \\
\\
{\bf percolate} through tree,     & c. of trees \\
{\bf diffuse} between its leaves  &     \\
\\
\hspace*{0.3cm} ?                 & thermodynamic depth     \\
\\
{\bf understand} system           &     \hspace*{0.3cm} ?     \\
\hline
\hline
\end{tabular}
\end{table*}

\section{APPLICATIONS}

To my knowledge, there have not yet been any real applications of the above ideas to non-trivial 
and realistic physical systems. The ``applications" listed below have to be viewed more as toy 
examples than anything else. But I firmly hope that real applications that deserve this name will 
come sooner or later.

\subsection{Complexities for the Quadratic Map [32,46]}

For the R-L symbol sequences generated by the quadratic map, we have different behavior in chaotic, 
periodic, intermittency, and Feigenbaum points. In the chaotic domain, we have also to distinguish 
between Misiurewicz (band-merging) points and typical chaotic points.

While all complexities (EMC, RLC, SC, FC) are small for periodic orbits, the EMC and the SC are 
infinite at Feigenbaum points. The reason is that there the block entropies $H_n$ diverge 
logarithmically. Indeed, one can show that for the Feigenbaum attractor the sequence $LL$ is 
forbidden, while the probabilities for the 3 other length-2 blocks are the same: 
$P(LR)=P(RL)=P(RR)=1/3$. Thus, $H_2 = \log 3$. For blocks of even length $2n$, with $n \geq 2$, one 
furthermore finds that $H_{2n} = H_n + \log 2$. Together with the monotonicity of $h_n$, this 
gives finally $h_n \sim const/n$ and $H_n \sim const \log n$. Thus the symbol sequences on 
the Feigenbaum attractor are so restricted that any continuation is unique with probability 1, 
but finding out the right continuation is very difficult. From time to time we have to go very 
far back in order to resolve an ambiguity \footnote{The same holds, by the way, also for 
quasiperiodic sequences and for Penrose tilings of the plane. There, the entropy increases 
logarithmically with the area. This is in my opinion the biggest obstacle against accepting the 
observed quasicrystals as perfect quasiperiodic lattices. While there would be no problem in a 
strict equilibrium state, the times necessary to resolve the constraints of a perfect quasilattice 
are astronomic.}.

\begin{figure}
\includegraphics[width=0.5\textwidth,angle=0]{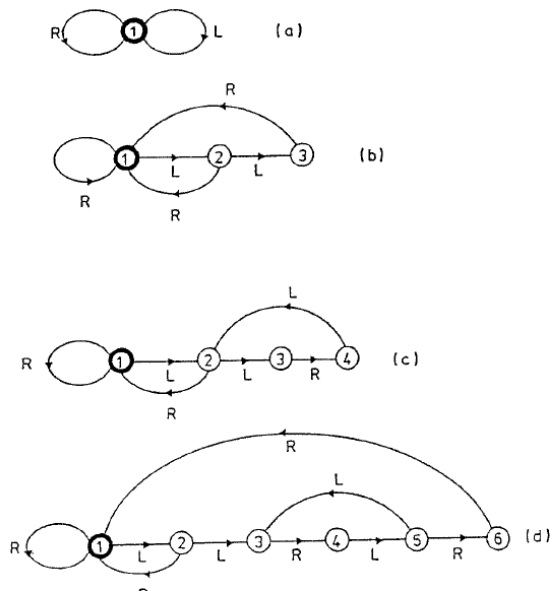}
\caption{Finite automata accepting the first approximations to the symbol sequence grammar for 
    the map $x' = 1.8 - x^2$. Graphs (a) - (d) accept all sequences which have correct blocks of 
    lengths 1, $\leq 3$, $\leq 4$, and $\leq 6$. The heavy node is the start.}
\end{figure}

For chaotic orbits, we have the problem (mentioned already in the analog context of sophistication) that for sequences depending on real parameters the forecasting complexity is infinite: when forecasting a very long sequence, it helps in using the control parameter a with ever increasing precision, leading to a divergent amount of work per symbol.

This does not apply to the EMC and to the SC. Block entropies should converge exponentially [47], 
so the EMC should be finite in general, as verified also numerically [46]. Also, there exists a 
simple algorithm for approximate grammars which accept all sequences containing no forbidden words 
of length $\leq n$ with any $n$ [48,46]. The graphs accepting symbol sequences of the map $x' = 1.8 
- x^2$ correctly up to length 1, 3, 4, and 6 are e.g. shown in Fig.~12. Except at Misiurewicz points, 
the size of these graphs diverges with $n$ (so that the RLC is infinite), but numerically the SC seems 
to stay finite. Thus one needs only a finite effort per symbol to check for grammatical correctness, 
for nearly all sequences with respect to the natural measure [46].

Exactly at intermittency points, maps with a parabolic critical point and with a quadratic tangency 
have the stationary distribution concentrated at the tangency point, and have thus trivial symbolic 
dynamics. But the SC can be shown to diverge logarithmically when an intermittency point is approached 
from below, again due to the divergence of the typical time scale.

\begin{figure}
\includegraphics[width=0.5\textwidth,angle=0]{fig-13}
\caption{Set complexity obtained with the natural invariant measure for the logistic map 
    $x' = a - x^2$ versus the control parameter $a$ (upper part). In the lower part, the bifurcation
    diagram is shown (from [46]).}
\end{figure}

A plot of the SC versus the parameter $a$ is given in Fig.~13. In order to locate better the various 
structures, we show below it the bifurcation diagram.

\subsection{Complexities of Grammars for Cellular Automata [39, 32]}

Wolfram [39] has studied the grammars of spatial strings $s_i,\; i\in Z$, generated by 1-dimensional
CA's after a finite number of iterations (the input string is taken as random). He finds that after 
any finite number of iterations the languages are always regular (this holds no longer if one goes 
to CA's in 2 dimensions, or to the strings after infinitely many iterations [49]).

One finds that for some rules the RLC increases very fast with the number of iterations. In many cases 
this corresponds to an actually observed intuitive complexity of the generated patterns, but for some 
rules (like, e.g., rules 32 or 160) the generated patterns seem rather trivial. In these latter cases, 
there is indeed a large difference between the RLC and the SC, the latter being very small [32]. 
Thus, the invariant measure is very unevenly distributed over the accepting deterministic graphs. 
Most of their parts are hardly ever used there during a scan of a typical sequence.

\subsection{Forecasting Complexities for Cellular Automata}

The only class of sequences for which we were able to compute finite forecasting complexities 
exactly were sequences generated by CA's after {\it a single iteration step}. Cellular automata with 
just one iteration are of course ridiculously simple systems, and one might expect very trivial 
results. But this is not at all so.

Assume there is a random stream of input bits, and at each time step one output bit is formed out 
of the last 3 input bits. The question one is asked is to predict as good as possible the probabilities
$p_i(0)$ and $p_i(1)$ for the i-th output bit to be 0 or 1, based only on the knowledge of previous 
output bits (in physics language, the input sequence is a ``hidden variable").

It is possible to give the optimal strategy for such an forecast [41]. It involves constructing a 
deterministic graph similar to those needed for regular languages. But now to each link is attached, 
in addition to the label $s$, a forecast $p(s)$. The FC is then given by the Shannon formula 
Eq.~(\ref{SC}), with the {\it grammar-}recognizing graph replaced by the graph producing the optimal
forecast.  While it is true that the FC is finite for all elementary CA's, the graphs are infinite 
for many rules [41].

Let me illustrate the method for rule 22. In this cellular automaton, the rule for updating the spins 
is that 001, 010, and 100 give ``1" in the next generation, while the other 5 neighborhoods give ``0". 
We write this as
\be
   s = F(t,t',t'')   
\ee 
with $F(0,0,0) = 0,\; F(0,0,1) = 1$, etc.

Assume that we have a random input string of zeroes and ones, called $t_n, n=0,1,2,\ldots$. Let us 
denote by $P_n(t,t'| S_{n-1})$ the probability ${\rm prob}(t_{n-1}=t, t_n=t')$, {\it conditioned}
on the first $n-1$ output spins to be $S_{n-1} = s_1, s_2, \ldots s_{n-1}$. When forecasting the first
output bit $s_1$, the only information we can use is that the probabilities $P_1(t,t')$ are equal to 
1/4 for each of the 4 combination of $t$ and $t'$. Let us denote by $p_n(s|S_{n-1})$ the forecasted 
probability for expecting the $n$-th output spin $s_n$ to be $s$, again conditioned on $S_{n-1}$.
One easily convinces oneself that
\be
   p_n(s|S_{n-1}) = 1/2  \sum_{t,t',t''} P_n(t,t'|S_{n-1}) \delta[s -F(t,t',t'')].    \label{4.2}
\ee
For $n=1$, this gives $p_1(0|.)=5/8,\; p_1(1|.) = 3/8$. After having predicted the next output by 
means of Eq.~(\ref{4.2}), we are told its actual value $s_n$ and have to update the probabilities 
for the input row. One finds
\bea
   P_{n+1} (t',t''| S_n) &=& [2p_n(s_n|S_{n-1})]^{-1} \sum_t P_n(t,t'|S_{n-1}) \times \nonumber \\
       &\times& \delta[s_n -F(t,t',t'')]. 
                 \label{4.3}
\eea
In this way, we alternate between the updating (\ref{4.3}) of the input probability, and the actual 
forecast (\ref{4.2}).

Each different value of the array $\{P_n(t,t''|S_{n-1}):\;\; (t,t') = (00)\ldots (11)\}$ 
corresponds to a node of the forecasting graph. Different nodes imply in general different forecasts 
in the next step\footnote{This is actually only true after minimizing the graph as described for 
the purely algorithmic case in ref.[33].}, while the same forecasts are obtained each time the 
same node is passed.

\begin{figure}
\includegraphics[width=0.5\textwidth,angle=0]{fig-14}
\caption{Part of the graph needed to forecast a sequence generated by CA rule 22 after 1 iteration [41].
    The heavy node is the start node. To each link are associated fixed forecasting probabilities 
    $p(s=1)$ and $p(s=0)=1-p(s=1)$, where $s$ is the next output symbol. In some of the nodes, we have 
    indicated the value of $p(s=1)$.}
\end{figure}

Part of the graph for rule 22 obtained in this way is given in fig.14. For more details, see ref.[41].

\subsection{Effective Measure Complexity for Cellular Automaton \# 22}

The most interesting application in view of possible self-generation of complexity is to the CA nr. 22. 
In this cellular automaton, the rule for spin updating is that 001, 010, and 100 give``1", while the 
other 5 neighborhoods give ``0". When starting from a single ``1", one gets a Pascal's triangle, and 
when starting from a random configuration one finds a pattern (Fig.~15) which at first sight does not 
look very interesting.

\begin{figure}
\includegraphics[width=0.5\textwidth,angle=0]{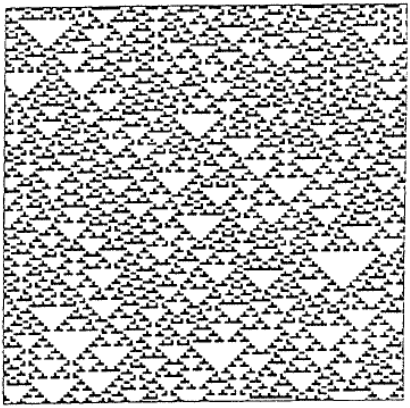}
\caption{Pattern generated by CA nr. 22 from a random start. Time runs again downwards, space horizontally.}
\end{figure}

\begin{figure}
\includegraphics[width=0.5\textwidth,angle=0]{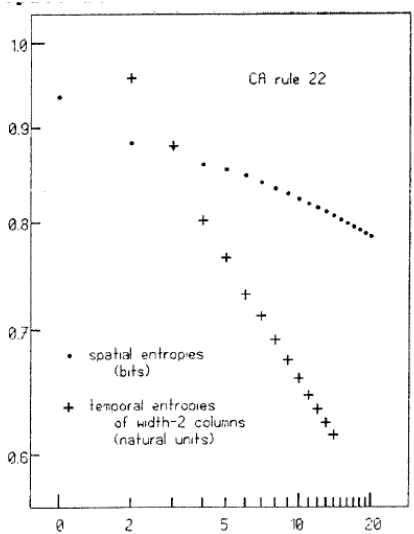}
\caption{Entropies $h_n$ for spatial sequences in bits (dots) and for time entropies in natural units
    (crosses) in the stationary state of CA 22, versus block length.}
\end{figure}

A closer inspection shows, however, that the block entropies for horizontal and for vertical blocks in 
the stationary state (reached after several thousand iterations) do not seem to increase linearly with 
the block length. Otherwise said, the differences $h_N$ seem to decrease like powers of $N$ (fig.16), 
such that both temporal and spatial strings have zero entropy in the stationary state. This is very 
surprising as the block entropies do diverge. Hence these strings are neither periodic nor fractal 
nor quasiperiodic nor random. Indeed, to my knowledge they are not like anything which other authors 
have encountered anywhere else.

The interesting aspect of this result is that it is very similar to an important aspect of life. Life 
is self-organizing in the sense that it leads to very special forms, i.e. from a wide basin of attraction 
it leads to a much narrower set of meaningful states. But this alone would not yet be surprising. The 
surprising aspect is that this attraction is not at all rigid. Although the ``attractor" is very small 
compared to full phase space, it is still huge and it therefore allows for a wide spectrum of behavior. 
It is exactly this which is shared by rule 22. Having zero entropy, the attractor is extremely 
constrained, but having divergent block entropies it is still vast.

\subsection{Entropy and Effective Measure Complexity of Written English}

I have looked at a number of other sequences whether they show similar long-range correlations, 
leading to a similarly small entropy. One does not seem to encounter this in symbol sequences 
generated by dynamical systems with few degrees of freedom. One does however encounter it to some 
degree in natural languages. In written English, e.g., the entropies $h_N$ decrease from 4.4 bits 
per character for $N=1$ to less than 1 bit for large $N$ [14].

A direct application of the defining equations (\ref{Hn}) - (\ref{hn}) to natural languages is 
unpractical due to the very long range of correlations. Alternative methods were devised already 
by Shannon [13]. In these, mutilated text was given to native speakers who could be assumed to be 
familiar with the grammar and probability measure of the text, but not with the text itself. The 
redundancy (and thus also the entropy) of the text was estimated from the percentage of letters which 
could be crossed out without impairing the understanding, and by the ability to complete letters of 
unfinished phrases.

\begin{figure}
\includegraphics[width=0.5\textwidth,angle=-1]{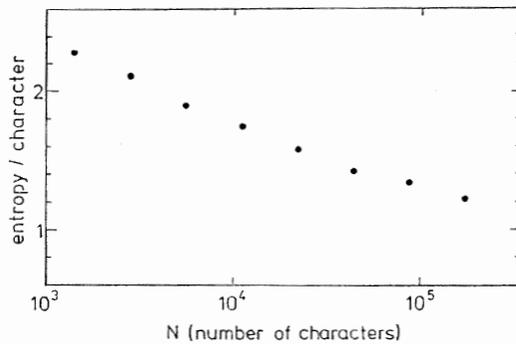}
\caption{Entropy estimates for written English based on the average lengths of repeated letter 
   sequences (from ref.[14]). On the horizontal axis is shown the total text length.}
\end{figure}

Objective entropy estimates were based in ref.[14] on the average length of sequences which repeat 
at least twice in a text of total length $N$. This length is given asymptotically by Eq.~(\ref{LZ}). 
Within N = $2\times 10^5$ letters, the average length of repetitions was $\approx 8$ letters. Since 
the average word length in English is $\approx$ 5.5 letters, one is then already sensitive to 
grammatical constraints between words. The decrease of the resulting entropy estimate (after taking 
into account leading corrections to Eq.~(\ref{LZ})) is shown in fig.17. We see a rather slow decrease 
-- indicating a large EMC as expected -- to a very small entropy value. We cannot indeed exclude 
that also there $h = 0$, with a power law convergence towards this limit\footnote{The latter seems
unlikely in view of the more recent analysis in arXiv:physics/0207023.}.

An attempt to analyze in a similar way DNA sequences failed due to extremely high repetition rates 
and non-stationarities. While the assumption that written English forms an ensemble with well-defined 
statistical properties seems reasonable, this is not so for DNA.

\vskip 1.cm

\noindent {\large \bf REFERENCES}

\vskip .3cm

\noindent 1. H. Haken, {\it Synergetics: an Introduction} (Springer, Berlin 1983)  \\
2. J. Guckenheimer and P. Holmes, {\it Non-Linear Oscillations, Dynamical Systems, 
  and Bifurcations of Vector Fields} (Springer, New York 1983)  \\
3. E.R. Berlekamp, J.H. Conway, and R.K. Guy, {\it Winning Ways for your 
   Mathematical Plays} (Academic Press, London 1983)  \\
4. M.V. Yakobson, Commun. Math. Phys. 81, 39 (1981)  \\
5. M. Feigenbaum, J. Stat. Phys. 19, 25 (1978); 21,669 (1979)  \\
6. Peitgen and P. Richter, {\it The Beauty of Fractals} (Springer, 1986)  \\
7. P. Collet and J.-P. Eckmann, {\it Iterated Maps on the Interval as Dynamical Systems} (Birkhauser, Basel 1980)  \\
8. A.N. Kolmogorov, ``Three Approaches to the Quantitative Definition of Information", Probl. of Inform. Theory 1, 3 (1965)  \\
9. G.J. Chaitin, Journal of the A.C.M. 22, 329 (1975)  \\
10. G.J. Chaitin, {\it Algorithmic Information Theory}, Cambridge University Press, 1987\\
11. C.E. Shannon and W. Weaver, {\it The Mathematical Theory of Communications} (Univ. of Illinois Press, Urbana, Ill., 1949)\\
R. Ash, {\it Information Theory} (Interscience Publishers, New York, 1965)\\
12. P. Grassberger, Phys. Lett. 128, 369 (1988)  \\
13. C.E. Shannon, Bell Syst. Tech. J. 30, 50 (1951)  \\
14. P. Grassberger, to appear in IEEE Trans. Inform.Th. (1989)  \\
15. S. Wagon, Mathem. Intell. 7, 65 (1985)  \\
16. J. Ziv and A. Lempel, IEEE Trans. Inform. Theory 24, 530 (1978)  \\
17. T.A. Welch, Computer 17, 8 (1984)  \\
18. G. Fahner and P. Grassberger, Complex Systems 1, 1093 (1987)  \\
19. A. Lempel and J. Ziv, IEEE Trans. Inform. Theory 22, 75 (1974)  \\
20. J. Rissanen, IEEE Trans. Inform. Theory 32, 526 (1986)  \\
21. G.J. Chaitin, ``Toward a Mathematical Definition of 'Life'", in 
    {\it The Maximum Entropy Principle}, R.D. Levine and M. Tribus, eds. (MIT Press, Cambridge, Mass., 1979)  \\
22. H. Simon, Proc. Am. Phil. Soc. 106, 467 (1962); \\
    H.A. Cecatto and B.A. Huberman, Physica Scripta 37, 145 (1988)  \\
23. D.R. Hofstatter, {\it Goedel, Escher, Bach: An Eternal Golden Braid} 
    (Vintage Books, New York 1980)  \\
24. S. Wolfram, Rev. Mod. Phys. 55, 601 (1983)  \\
25. S. Wolfram, {\it Glider Gun Gidelines}, Princetown preprint (1985)  \\
26. M.H. van Emden, {\it An Analysis of Complexity} (Mathematical Centre Tracts, Amsterdam 1975); \\
    J. Rothstein, in ``The Maximum Entropy Formalism", R.D. Levine 
    and M. Tribus, Eds. (Cambridge Univ. Press, 1979) \\
27. M. Gardner, Sci. Amer. 224, issue 2, p.112 (Feb. 1971)  \\
28. P. Grassberger, J. Stat. Phys. 45, 27 (1986)  \\
29. R. Dawkins, {\it The Blind Watchmaker} (Norton, New York, 1987) \\
30. S. Lloyd and H. Pagels, Annals of Phys. 188, 186 (1988)  \\
31. H. Atlan, Physica Scripta 36, 563 (1987)  \\
32. P. Grassberger, Int. J. Theor. Phys. 25, 907 (1986)  \\
33. J.E. Hopcroft and J.D. Ullman, {\it Introduction to Automata Theory, Languages, and Computation} (Addison-Wesley, 1979)\\
34. D. Ruelle, {\it Thermodynamic Formalism} (Addison-Wesley, Reading, Mass., 1978)  \\
35. C.H. Bennett, in ``Emerging Syntheses in Science", D.Pines ed., 1985  \\
36. S. Wolfram, {\it Random Sequence Generation by Cellular Automata}, to appear in Adv. Appl. Math.  \\
37. M. Koppel and H. Atlan, {\it Program-Length Complexity, Sophistication, and Induction}, preprint (1987)  \\
38. R.W. Hamming, {\it Coding Theory and Information Theory} (Prentice- Hall, Englewood Cliffs, 1980)  \\
39. S. Wolfram, Commun. Math. Phys. 96, 15 (1984)  \\
40. K. Lindgren and M. Nordahl, Chalmers Univ. preprint (1988)  \\
41. D. Zambella and P. Grassberger, Complex Systems 2, 269 (1988)  \\
42. J.P. Crutchfield and K. Young, Phys. Rev. Lett. 63, 105 (1989)  \\
43. T. Hogg and B.A. Huberman, Physica D 22, 376 (1986)  \\
44. C.P. Bachas and B.A. Huberman, Phys. Rev. Lett. 57, 1965 (1986); J. Phys. A 20, 4995 (1987)  \\
45. C.P. Bachas and W.F. Wolff, J. Phys. A 20, L39 (1987)  \\
46. P. Grassberger, Z. Naturforsch. 43a, 671 (1988)  \\
47. G. Gy\"orgyi and P. Szepfalusy, Phys. Rev. A31, 3477 (1985)  \\
48. F. Hofbauer, Israel J. Math. 34, 213 (1979); 38, 107 (1981)  \\
49. M. Nordahl, Chalmers Univ. thesis (1988)  \\

\end{document}